# Exploring Chaotic Motion of a Particle in the Centre of a Galaxy with a Prolate Halo

**Uditi Nag\***, Yeasin Ali[a], Suparna Roychowdhury[a]

*\*School of Physics and Astronomy, Cardiff University, Cardiff, CF24 3AA, UK*
*[a]Department of Physics, St. Xavier's College, 30 Park Street, Kolkata-16, India*

\*Corresponding Author E-mail: uditi.work08@gmail.com

## Abstract

The majority of galaxies are known to have supermassive black holes (SMBHs) at their core, which have a tremendous gravitational pull on the objects around them. When embedded within extended matter distributions like prolate, shell-like halos, they give rise to complex gravitational fields that often drive nearby particles into chaotic orbits. The inherently nonlinear nature of such motion, shaped by general relativity, makes direct analysis highly challenging. To overcome this, pseudo-Newtonian potentials are used to approximate relativistic effects within a Newtonian framework. In this study, we model the central SMBH using the Artemova–Björnsson–Novikov (1996) potential to mimic the rotational effects of a Kerr-like BH. The surrounding prolate halo is treated as an axisymmetric, shell-like mass distribution, represented through a multipole expansion including dipole, quadrupole components. Poincaré sections reveal how the SMBH–halo system drives both order and chaos, with the SMBH's spin modulating dynamics by enhancing or suppressing chaos depending on its direction and magnitude.

## 1. Introduction

The central regions of galaxies provide an exceptional environment for exploring the dynamics of compact objects and dense stellar systems. Observations strongly indicate the coexistence of supermassive black holes (SMBHs) and nuclear star clusters in these regions [1]. Bulges may appear in classical or boxy–peanut morphologies, while nuclear clusters typically range between $10^5$ and $10^8 M_\odot$ [1]. SMBHs are often encircled by accretion disks and extended matter halos, both of which strongly influence nearby stellar orbits [1].

To model such environments, analytical potentials such as the Hernquist [2] and Navarro–Frenk–White (NFW) [3] profiles are widely used, although incorporating features like shells, rings, or asymmetric halos remains challenging. Since the SMBH mass vastly exceeds that of individual stars, stars are often treated as test particles, with gravitational interactions between them neglected [4]. A common simplification employs a monopole potential for the compact object, supplemented by higher-order multipolar terms (dipole, quadrupole, octupole) to capture halo asymmetries [5–11].

While general relativity provides the most accurate description of supermassive black hole (SMBH) environments, solving Einstein's equations for realistic galactic systems is computationally demanding. Pseudo-potential approaches, therefore, offer an efficient alternative. The Artemova–Björnsson–Novikov (ABN) potential [7], along with related formulations, models Kerr-like compact objects while preserving analytical simplicity. In this study, we consider a test particle moving in a combined potential consisting of a central pseudo-Newtonian monopole representing the SMBH and multipolar contributions describing the galactic halo.

Such systems are inherently non-integrable. A spherically symmetric black hole potential alone does not produce chaos, but even small non-central perturbations can trigger complex dynamics [11]. Previous studies have shown that quadrupolar terms do not generate chaos on their own, whereas octupolar contributions do [11]. Moreover, oblate halos generally enhance chaotic motion, while prolate halos reduce it [11]. Comparisons between Newtonian and relativistic frameworks reveal that relativistic corrections typically improve orbital stability, though the effect depends on spin and halo structure [6,8,10]. Importantly, spin often weakens chaotic behavior in dipolar or quadrupolar halos [6,10].

The objective of this work is to explore how halo multipolar moments—dipole and quadrupole—interact with SMBH spin to shape particle dynamics. Section 2 introduces the mathematical framework, Section 3 presents orbital structures and Poincaré maps in Newtonian and special relativistic regimes, and Section 4 discusses the main results.

## 2. Methods

### 2.1. Mathematical Formulation

The core mass distribution of galaxies containing a central compact object can be efficiently modeled using a multipole expansion (Binney & Tremaine, 2008) [5]. In our formulation, terms up to the quadrupole order are included. The monopole describes the central SMBH, the quadrupole ($Q < 0$) models a prolate halo shell, and the dipole (D) term accounts for asymmetries in the halo. To represent the central compact object more realistically, we replace the Newtonian monopole with the Artemova–Björnsson–Novikov (ABN) pseudo-potential, which mimics the gravitational field of a Kerr-type object (Artemova et al., 1996) [7]. The ABN potential in cylindrical coordinates (ρ, z) can be written as [6,10]:

$$\Phi_{ABN}(\rho, z) = -\frac{1}{r_1(\beta-1)} \times \left[\frac{(\rho^2+z^2)^{\frac{\beta-1}{2}}}{\left(\sqrt{(\rho^2+z^2)}-r_1\right)^{(\beta-1)} - 1}\right]$$

The total gravitational potential of the system is [10,11]:

$$\Phi_g = \Phi_{ABN}(\rho, z) + Dz + \left(\frac{-Q}{2}\right)(2z^2 - \rho^2) \qquad (1)$$

Including the centrifugal term from conserved angular momentum L, the effective potential is:

$$V(\rho, z) = \Phi_g + \frac{L^2}{2\rho^2}$$

**Newtonian Dynamics:**

So, the equations of motion for Newtonian dynamics can be written as [8,11], as shown in the box below-

$$\dot{\rho} = p_\rho \qquad (2a)$$

$$\dot{p}_\rho = -\frac{\partial V}{\partial \rho} \qquad (2b)$$

$$\dot{z} = p_z \qquad (2c)$$

$$\dot{p}_z = -\frac{\partial V}{\partial z} \qquad (2d)$$

So applying these energy and angular momentum conservation conditions for this system, we can find the region where the test particle is restricted to move. That region is defined as

$$E_{mec} = \frac{p_\rho^2 + p_z^2}{2} + \frac{L^2}{2\rho^2} + \Phi_g, \text{ where } E = \sqrt{1 + 2E_{mec}} \qquad (3)$$

**Relativistic Dynamics:**

Maintaining the non-relativistic definitions of $p_\rho$ and $p_z$ unchanged for the sake of comparison, the equations of motion take the form [8,11], as shown

$$\dot{\rho} = p_\rho \qquad (4a)$$

$$\dot{p}_\rho = \frac{1}{\Phi_g - E}\left[\frac{\partial \Phi_g}{\partial \rho}(1 - p_\rho^2) - \frac{\partial \Phi_g}{\partial z}p_z p_\rho - \frac{L^2}{(E - \Phi_g)\rho^3}\right] \qquad (4b)$$

$$\dot{z} = p_z \qquad (4c)$$

$$\dot{p}_z = \frac{1}{\Phi_g - E}\left[\frac{\partial \Phi_g}{\partial z}(1 - p_z^2) - \frac{\partial \Phi_g}{\partial \rho}p_z p_\rho\right] \qquad (4d)$$

For relativistic dynamics, the test particle is restricted to move in region defined as

$$p_\rho^2 + p_z^2 + \frac{L^2}{(E-\Phi_g)^2} + \frac{1}{(E-\Phi_g)^2} = 1 \qquad (5)$$

## 2.2. Poincaré Section

Here, the Poincaré section method is utilized to investigate the dynamical properties of the system. The test particle's motion is confined to the equatorial plane (z=0) within the four-dimensional phase space $(\rho, z, p_\rho, p_z)$, constrained by conserved energy (E) and angular momentum (L). Trajectories are evolved with MATLAB (ode45) in the equatorial plane ($z = 0$, $p_z > 0$) with angular momentum, L = 4.2, and energy, E = 0.976. These values were chosen as a direct extension of [10], but applied here to a prolate halo. Equations are integrated while varying spin (a) and halo ratio (P). Quadrupole elongation and spin govern the transition from regular to chaotic motion in the Poincaré maps. The Poincaré section is obtained as a 2D slice ($\rho, p_\rho$) at each crossing of the $z = 0$ plane with $p_z > 0$. This projection clearly distinguishes orbital behaviors: smooth invariant curves indicate regular motion, quasi-periodic (sticky) orbits align with KAM-like tori, and chaotic orbits scatter across the plane. As chaos strengthens, regular orbit islands disintegrate into scattered points.

## 3. Results and Discussion

The chaotic dynamics of the system are influenced by both the halo ratio $P = \frac{D}{Q}$ and the spin parameter a. As | P | increases, the dipole contribution 'D' grows, enhancing chaos, consistent with previous studies on multipolar galactic potentials [6,10,11]. Increasing the spin parameter, however, suppresses chaos. For low spin (a = 0.1) and moderate | P |= 3.0, Newtonian and relativistic formulations display nearly identical chaotic behavior (Figure 2, plots a and b), with differences that are subtle and require sensitive indicators such as the maximum Lyapunov exponent.

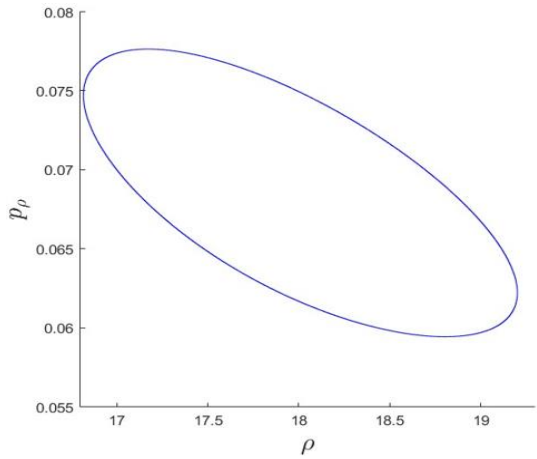
(a) Regular Poincaré Section

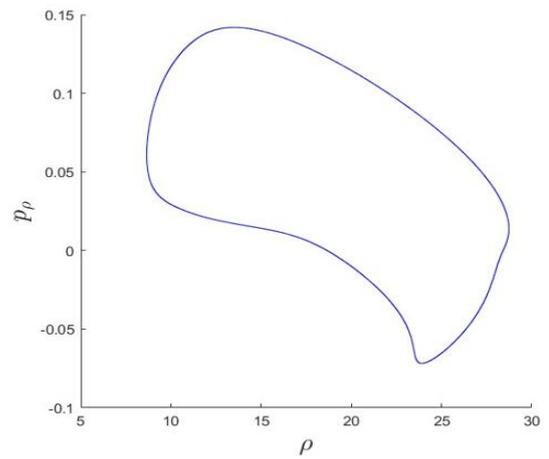
(b) Sticky Poincaré Section

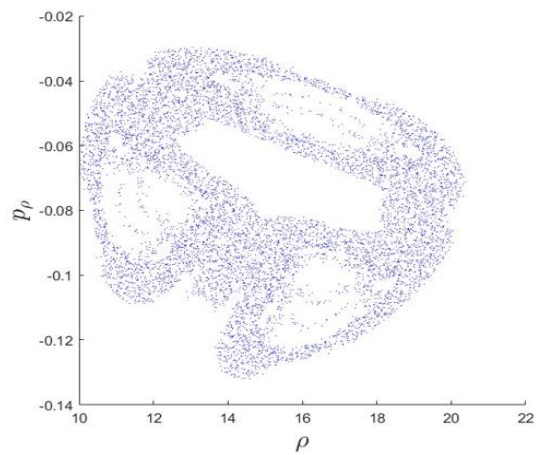
(c) Chaotic Poincaré Section

Figure 1: (a)-(c) shows the different types of Poincaré sections: regular, sticky, and chaotic, respectively. For these trajectories, $P = \frac{D}{Q} = -300$, is used. Common parameters: $E = 0.976, L = 4.2, Q = -1 \times 10^{-6}$.

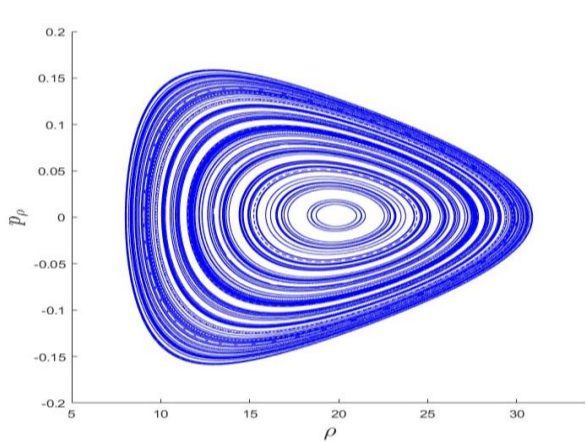
(a) Newtonian; a=0.1, P=-3

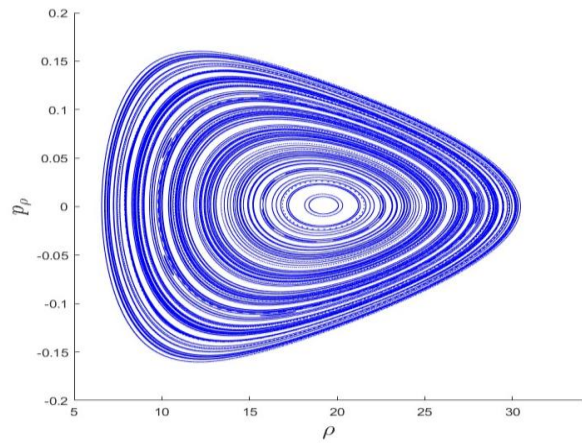
(b) Relativistic; a=0.1, P=-3

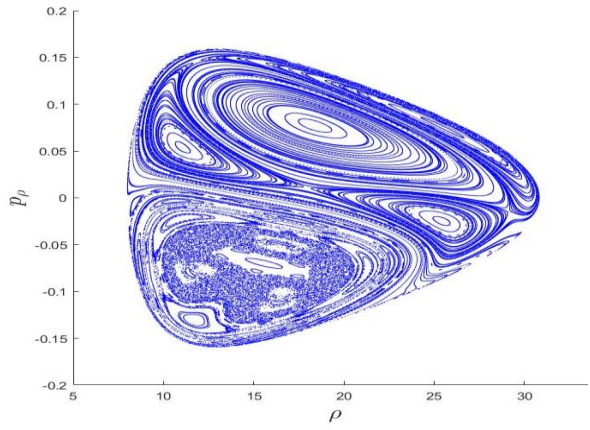
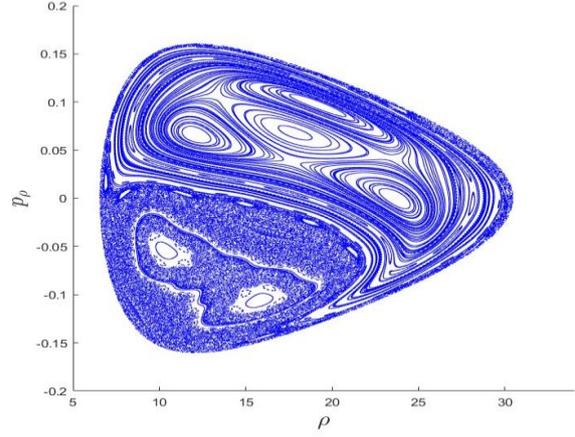

(c) Newtonian; a=0.1, P=-300    (d) Relativistic; a=0.1, P=-300

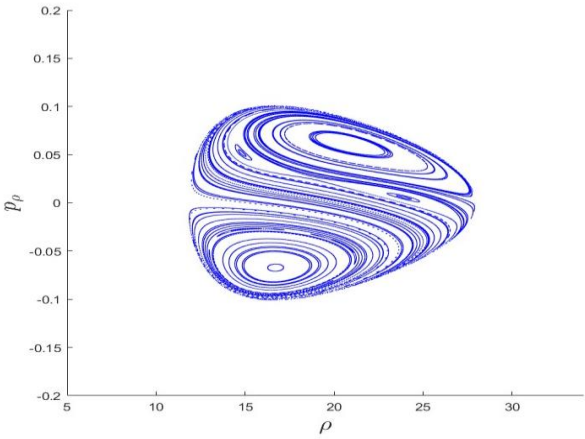
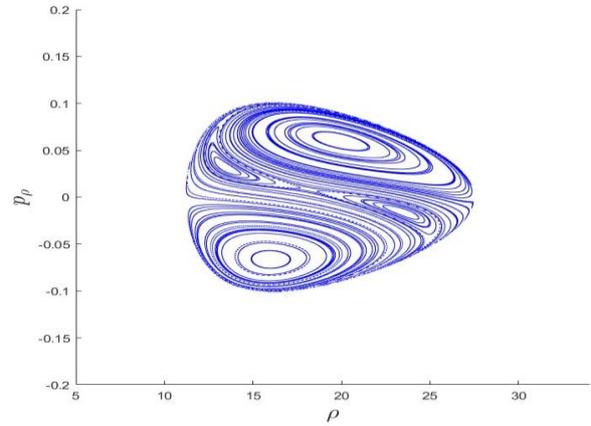

(e) Newtonian; a=0.9, P=-300    (f) Relativistic; a=0.9, P=-300

Figure 2: Poincaré sections for Newtonian (left) and relativistic (right) cases. Fixed parameters: $E = 0.976, L = 4.2$. Spin parameter a and halo ratio, $P = \frac{D}{Q}$, are varied, with $Q = -1 \times 10^{-6}$ in all simulations.

As | P | increases further (Figure 2, plots c and d), relativistic dynamics exhibit stronger chaos than Newtonian. At higher spin ($a = 0.9$), chaotic behavior in both formulations becomes comparable (Figure 2, plots e and f), indicating that higher spin systematically reduces chaos. Overall, these results demonstrate an inverse relationship between spin and chaos and highlight the need for advanced chaos indicators to resolve subtle differences between Newtonian and relativistic systems. We now use better indicators of chaos to study such subtle differences in chaos.

The Maximum Lyapunov Exponent (MLE) measures dynamical instability through the exponential divergence of nearby trajectories. As a coordinate-independent chaos indicator, it quantifies what we observe in Poincaré sections, reveals subtle instabilities, and enables direct comparison across parameters. In our Newtonian model, MLE shows how | P | and spin parameter, (a), govern dynamical stability. The analysis uncovers two dominant stabilizing mechanisms. First, (as in Figure 3) the dipole-to-quadrupole ratio, |P|, shows that as |P| increases, chaos increases, as seen from the MLE. Second, (as in Figure 4), the spin parameter (a) exhibits an equally strong effect, with chaos decreasing as spin increases.

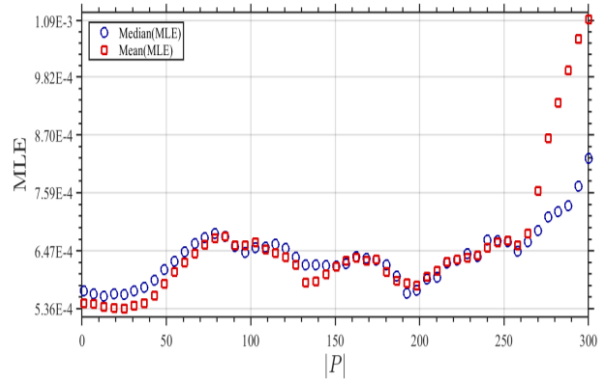

Fig 3: MLE decreases with increasing radial momentum |P|

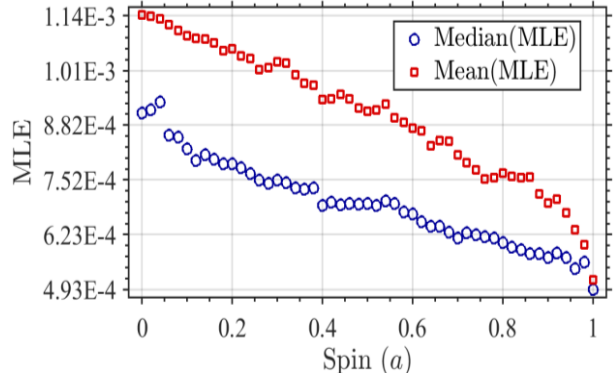

Fig 4: Stability increases with spin parameter $a$, as shown by the decreasing MLE.

The MLE findings reaffirm our observations from the Poincaré sections: as |P| increases, chaos increases, whereas higher spin values (a) lead to a decrease in chaos.

## 4. Conclusion

A quadrupole-truncated multipole expansion of the galaxy potential describes non-spherical galaxies and encapsulates important structural asymmetries like bars and spiral arms. The model enhances the description of stellar orbital and gas dynamics as well as the rich gravitational effects arising from galaxy mergers and tidal encounters. In galactic nuclei dominated by a supermassive black hole's potential, the quadrupole expansion enables precise simulations of quasi-Keplerian orbits and their long-term secular perturbations due to stellar encounters.

5. **References**

1. Genzel R, Eisenhauer F, Gillessen S. The Galactic Center massive black hole and nuclear star cluster. *Rev Mod Phys*. 2010;82(4):3121–95.
2. Hernquist L. An analytical model for spherical galaxies and bulges. *Astrophys J*. 1990;356:359–64.
3. Navarro JF, Frenk CS, White SDM. The structure of cold dark matter halos. *Astrophys J*. 1996;462:563–75.
4. Alexander, T. (2017). Stellar Dynamics and Stellar Phenomena Near a Massive Black Hole. Annual Review of Astronomy and Astrophysics, 55(1), 17–57. https://doi.org/10.1146/annurev-astro-091916-055306
5. Binney J, Tremaine S. *Galactic Dynamics*. 2nd ed. Princeton: Princeton University Press; 2008.
6. Nag S, Sinha S, Ananda DB, Das TK. Influence of the black hole spin on the chaotic particle dynamics within a dipolar halo. *Astrophys Space Sci*. 2017;362(4):92.
7. Artemova IV, Björnsson G, Novikov ID. Modified Newtonian potentials for a rotating black hole. *Astrophys J*. 1996;461:565–72.
8. Guéron E, Letelier PS. Chaos in pseudo-Newtonian black holes with halos. *Astron Astrophys*. 2001;368(2):716–20.
9. Wang Y, Wu X. Dynamics of particles around a pseudo-Newtonian Kerr black hole with halos. *Chin Phys B*. 2012;21(5):050504.
10. Ali, Y., & Roychowdhury, S. (2024). Chaotic dynamics in a galactic multipolar halo with a compact primary. *Physical Review E*, 110(6), 064202.
11. Vieira WM, Letelier PS. Relativistic and Newtonian core–shell models: Analytical and numerical results. *Astrophys J*. 1999;513(1):383–400.